\documentclass[preprint,12pt]{elsarticle}
\usepackage{amsmath,amssymb}
\usepackage{changes}
\usepackage{lipsum}
\definechangesauthor[name={XiaoRenjie}, color=yellow]{xrj}
\usepackage{algpseudocode}
\usepackage{tabularx}
\usepackage{graphicx}
\usepackage[ruled,vlined,linesnumbered]{algorithm2e}
\usepackage{subfigure} 
\usepackage{amssymb}
\usepackage{amsthm}
\usepackage{amsmath}
\usepackage{hyperref}
\usepackage{caption}
\usepackage{subcaption}
\usepackage{colortbl}
\usepackage{overpic}
\usepackage{diagbox}
\usepackage{booktabs}
\usepackage{array}
\usepackage{bm}
\usepackage{multirow}
\usepackage{booktabs}
\usepackage{textcomp}
\usepackage{wrapfig}
\usepackage[shortlabels]{enumitem}
\usepackage[title]{appendix}
\usepackage{tabularray}

\journal{Journal of Computational Physics}

\begin{document}

\begin{frontmatter}

\title{
Flow Field Reconstruction via Voronoi-Enhanced Physics-Informed Neural Networks with End-to-End Sensor Placement Optimization}

\author[inst1,inst2]{Renjie Xiao}
\author[inst1,inst2]{Bingteng Sun\texorpdfstring{\corref{correspondingauthor}}{}}\ead{sunbingteng@iet.cn}
\author[inst1,inst2]{Yiling Chen}
\author[inst6]{Lin Lu}
\author[inst1,inst2,inst3,inst4,inst5]{Qiang Du\texorpdfstring{\corref{correspondingauthor}}}
\author[inst1,inst2]{Junqiang Zhu}

\affiliation[inst1]{
    organization={Advanced Gas Turbine Laboratory, Institute of Engineering Thermophysics, Chinese Academy of Sciences}, 
    city={Beijing},
    postcode={100190}, 
    country={China}}

\affiliation[inst2]{
    organization={National Key Laboratory of Science and Technology on Advanced Light-duty Gas-turbine}, 
    city={Beijing},
    postcode={100190}, 
    country={China}}

\affiliation[inst3]{
    organization={University of Chinese Academy of Sciences},
    city={Beijing},
    postcode={100190}, 
    country={China}}

\affiliation[inst4]{
    Qingdao Institute of Aeronautical Technology}

\affiliation[inst5]{
    Nanjing Future Energy System Research Institute, Nanjing, Jiangsu 211135, China}

\affiliation[inst6]{
    School of Computer Science and Technology, Shandong University, Qingdao, Shandong, 266237, China}

\cortext[correspondingauthor]{Correspondingauthor}

\begin{abstract}
High-fidelity flow field reconstruction is a crucial research focus in fluid dynamics, yet it faces prominent challenges due to the sparsity, spatiotemporal incompleteness of measured sensor data and the frequent failure of pre-deployed measurement points, which invalidates pre-trained reconstruction models.
Physics-informed neural networks (PINNs) have been applied to field reconstruction with imperfect data by incorporating physical principles, thus reducing reliance on massive labeled sensor data. 
However, sensor placement optimization, a key factor for PINN performance improvement, has been largely overlooked in existing research.
Therefore, developing an intelligent algorithm for sensor placement optimization that enhances the robustness of PINNs against specific measurement point failures is of great practical significance.
In this study, a novel PINN with Voronoi enhanced Sensor Optimization (VSOPINN) is proposed, which innovatively realizes the differentiable construction of soft Voronoi diagrams for sparse sensor data rasterization, the end-to-end fusion of centroidal Voronoi tessellation (CVT) and PINNs for adaptive sensor placement optimization, and the unified sensor layout optimization for multi-condition flow field reconstruction via a shared encoder-multi-decoder architecture.
%
The effectiveness of VSOPINN in flow field inference across diverse geometric solution domains is validated through three typical cases: the lid-driven cavity flow, vascular flow, and annular rotating flow problems.
Experimental results show that the proposed VSOPINN not only significantly improves the test accuracy of field reconstruction for different Reynolds numbers but also adaptively learns the optimal sensor layout. Moreover, it exhibits remarkable robustness against the failure of partial measurement points.
This study deepens the understanding of the intrinsic relationship between sensor placement and reconstruction precision in PINN-based field reconstruction.

\end{abstract}


\end{frontmatter}


\section{Introduction}



Fluid physics is typically investigated through numerical simulations or experimental analyses of canonical flows. However, both approaches suffer from inherent limitations. Direct Numerical Simulation (DNS) \cite{Huang1995} and Large Eddy Simulation (LES) \cite{CHUNG2009} yield high-fidelity solutions to the Navier–Stokes equations but incur prohibitive computational costs, while the Reynolds-Averaged Navier–Stokes (RANS) method fails to resolve fine-scale flow features \cite{Carlberg2019}. Concurrently, experimental flow visualization and measurement demand substantial resource investment \cite{Deng2019}. Consequently, flow field reconstruction methods have emerged as a viable alternative, leveraging sparse sensor measurements to estimate the full flow field \cite{Callaham2019}.
Mathematically, this problem boils down to identifying the mapping $G$ between the point measurements $y(\boldsymbol{x}_c,t)\in\mathbb{R}^p$ and the complete flow field $\boldsymbol{U}(\boldsymbol{x},t)\in\mathbb{R}^n$, where $p$ denotes the number of sensors, $\boldsymbol{x}_c$ their spatial locations, and $n$ the dimensionality of the flow field. Given that the measurement operator $H$ is highly likely to be ill-posed and thus non-invertible, data-driven methods have been developed to estimate the operator $G$ from data\cite{Manohar2018}.

According to Dubois et al.\cite{Dubois2022}, flow field reconstruction methods are primarily classified into three major categories: direct reconstruction methods (Everson and Sirovich\cite{Everson1995}; Willcox\cite{Willcox2006}; Boisson and Dubrulle\cite{Boisson2011}; Callaham et al.\cite{Callaham2019}; BuiThanh et al.\cite{BuiThanh2004}), which directly learn the mapping relationship between measured data and flow fields via supervised learning; regression reconstruction methods (Sirovich\cite{Sirovich1987}; Erichson et al.\cite{Erichson2020}; Xu Shengfeng et al.\cite{XuShengFeng2023}), which represent flow fields as linear combinations of reference modes based on optimization problems and solve for the optimal combination coefficients using limited measurements; and data assimilation methods (Mons et al.\cite{Mons2016}; Colburn et al.\cite{COLBURN2011}; Zauner et al.\cite{Zauner2022}), which integrate dynamic model predictions with real-time measured data and revise the predicted results through an update mechanism to improve accuracy. Among these approaches, regression reconstruction methods based on machine learning and deep learning have been extensively applied in recent years \cite{Brunton2020}\cite{Shen2024}, benefiting from the rapid development of computer science and technology, with super-resolution reconstruction emerging as a key research direction\cite{Fukami2021}.
However, deep learning-based regression reconstruction methods for flow field reconstruction suffer from heavy reliance on large-scale high-quality labeled sensor data and poor generalization performance in data-sparse scenarios, making them hard to adapt to practical engineering with limited measurement data.

Notably, PINNs methods also fall into the category of regression reconstruction methods but represent a pivotal advancement in solving complex fluid governing differential equations by fusing deep learning with physical laws. NSFnet~\cite{Jin2021} introduces Navier-Stokes Flow Nets (NSFnets), a PINN variant tailored for incompressible laminar and turbulent flow simulations that embeds governing equations directly to reduce labeled data dependence. Subsequent work like PINN-SR~\cite{Chen2021} enhances approximation precision by integrating deep neural networks for representation learning and sparse regression. Further innovations include the meshfree, unsupervised Meta-Auto-Decoder~\cite{https://doi.org/10.48550/arxiv.2111.08823}, which uses meta-learning to encode PDE parameters as latent vectors for rapid model adaptation; Bai et al.\cite{Bai2020}, who extend PINNs to label-free cylinder flow simulation via continuum and constitutive formulations to enable flow data assimilation; and PhyGeonet~\cite{Gao2021}, which develops a physics-constrained CNN architecture for irregular domain learning by morphing complex meshes into square domains without labeled data.
However, the current PINN method performs poorly in solving highly nonlinear systems such as turbulent fields or three-dimensional scenes, so PINN at this stage should not only be treated as a Computational Fluid Dynamics (CFD) method, but also paid more attention to its advantages in processing sparse data, and PINN could be treated as a regression reconstruction method.
For indoor environments, Wei et al.\cite{Wei2023} adopted the PINN algorithm to reconstruct the two-dimensional isothermal indoor airflow field using sparse CFD data, which exhibits superior performance over traditional data-driven methods. In outdoor wind environments, Rui et al.\cite{Rui2023} reconstructed the three-dimensional flow field around an isolated building with sparse sensor data obtained from wind tunnel experiments, demonstrating that this method holds broad application prospects in the field of wind engineering.
These studies provide encouraging insights that the PINN can serve as a promising data inversion method.

While PINNs excel in sparse data processing, the fidelity and efficiency of all aforementioned reconstruction methods, including PINNs, are fundamentally constrained by the spatial distribution of sensor points. Regardless of the approach (direct reconstruction, regression, or data assimilation), sensor placement is a critical factor that dictates the accuracy and computational efficiency of the final reconstruction.
Various strategies have been proposed to optimize the location of sensors, i.e., sensor placement optimization methods.
Deng et al.\cite{Deng2021} focused on the optimization strategy based on a DNN for turbulent flow recovery within the data assimilation framework of the ensemble Kalman filter (EnKF). Cai et al.\cite{Cai2021} proposed an optimization method based on the PINN framework that selects sensor locations by evaluating residuals of governing equations. Sharma et al.\cite{Sharma2024} systematically generated 80 distinct sensor configurations for a 2-dimensional stenosis hemodynamics problem and demonstrated that the accuracy of flow-field predictions is notably more sensitive to sensors located close to the stenosis and inlet. 
However, these existing PINN-associated sensor optimization methods are either empirically designed without adaptive coupling to the PINN model’s training update or lack effective geometric adaptability for complex irregular flow domains, leading to suboptimal sensor layouts and poor robustness in practical engineering with sparse and incomplete measurements.
We can see that the existing optimization methods are generally decoupled from the field reconstruction algorithm, so the optimization of sensor placements does not adaptively match the updates of the reconstruction model\cite{Xie2024}. In particular, the optimization algorithm combined with the PINN method is basically empirical \cite{Zhu2022}.
In addition, existing regression reconstruction methods face a practical application challenge, pre-deployed sensors are often randomly damaged due to environmental factors and other uncertainties \cite{SBrooks2018}. This leads to the failure of reconstruction models designed for a fixed number of measurement points due to inconsistent input, resulting in the waste of training costs.

In response to the aforementioned challenges, we propose a model that incorporates the sparse sensor placement optimization method into a PINN, called VSOPINN. 
In summary, this work makes three key contributions to flow field reconstruction with sparse sensors: 
\begin{itemize}
\item[$\bullet$] We propose a differentiable soft Voronoi diagram construction method that converts unstructured sensor data into grid-based representations for CNN feature extraction;
\item[$\bullet$] We realize end-to-end fusion of CVT and PINNs for adaptive sensor placement optimization, which enables the model to autonomously identify high-information-entropy regions in flow fields; 
\item[$\bullet$] We design a shared encoder-multi-decoder architecture for unified sensor layout optimization across multi-condition flow fields, improving the generalization ability of the model under varying Reynolds numbers. 
\end{itemize}

The structure of this paper is organized as follows: 
In Section~\ref{sec:methods}, after providing a brief overview of PINNs, we provide a detailed introduction to the VSOPINN model, including the Voronoi image generation method, Voronoi image Encoder, PDE solver Decoder, and multi-VSOPINN model.
In Section~\ref{sec:results}, we conduct a series of systematic numerical experiments to validate the efficacy and robustness of the proposed VSOPINN framework and evaluate its performance across three key flow cases.
Additionally, within these cases, we compare the performance of multiple model configurations, including baseline methods and VSOPINN variants, across different types of PDE problems.
In Section~\ref{sec:discussion}, we summarize the findings and describes future research directions.
\section{Methods}
\label{sec:methods}

\subsection{Physics-informed neural networks}
\label{sec:PINN}

Physics-informed neural networks (PINNs) embed the governing partial
differential equations (PDEs) into the training of a neural surrogate, so
that the learned model is constrained to satisfy the underlying physics in
addition to fitting the available data. Let $\boldsymbol{x} \in \Omega
\subset \mathbb{R}^d$ denote the spatiotemporal coordinates, and let $\boldsymbol{q}(\boldsymbol{x})$ collect the flow
variables of interest, e.g.\ velocity components and pressure. The dynamics
of the system are modeled by a nonlinear differential operator
$\mathcal{N}$ and suitable boundary operators $\mathcal{B}$,
\begin{equation}
  \begin{aligned}
    \mathcal{N}\bigl(\boldsymbol{q}(\boldsymbol{x});\,\boldsymbol{\mu}\bigr)
    &= \boldsymbol{0}, && \boldsymbol{x} \in \Omega,\\
    \mathcal{B}\bigl(\boldsymbol{q}(\boldsymbol{x});\,\boldsymbol{\mu}\bigr)
    &= \boldsymbol{0}, && \boldsymbol{x} \in \partial\Omega,
  \end{aligned}
  \label{eq:pde_bc_general}
\end{equation}
where $\boldsymbol{\mu}$ denotes physical parameters e.g.\ viscosity,
density, Reynolds number.

Instead of discretizing the PDE system~\eqref{eq:pde_bc_general} on a fixed mesh, PINNs
approximate the unknown field by a neural network
$\boldsymbol{q}_\theta(\boldsymbol{x})$ with trainable parameters
$\theta$. All spatial and temporal derivatives required by $\mathcal{N}$
and $\mathcal{B}$ are obtained via automatic differentiation with respect
to the inputs $\boldsymbol{x}$, which avoids numerical differencing and
keeps the model fully differentiable with respect to both $\theta$ and
$\boldsymbol{\mu}$.

Given this surrogate, we define the PDE residual and boundary residual as
\begin{equation}
  \begin{aligned}
    \boldsymbol{r}_\theta(\boldsymbol{x})
    &=
    \mathcal{N}\bigl(\boldsymbol{q}_\theta(\boldsymbol{x});\,\boldsymbol{\mu}\bigr),
    && \boldsymbol{x} \in \Omega,\\
    \boldsymbol{r}^{\mathrm{bc}}_\theta(\boldsymbol{x})
    &=
    \mathcal{B}\bigl(\boldsymbol{q}_\theta(\boldsymbol{x});\,\boldsymbol{\mu}\bigr),
    && \boldsymbol{x} \in \partial\Omega,
  \end{aligned}
  \label{eq:pde_bc_residual}
\end{equation}
The network parameters are learned by minimizing a composite loss function
that combines data and physics residuals. Let
$\{\boldsymbol{x}_i^{d}, \boldsymbol{q}_i^{\mathrm{obs}}\}_{i=1}^{N_d}$ be
sensor measurements, $\{\boldsymbol{x}_j^{f}\}_{j=1}^{N_f}$ the collocation
points inside the domain, and $\{\boldsymbol{x}_k^{b}\}_{k=1}^{N_b}$ points
on the boundary. A typical PINN loss reads
\begin{equation}
  \mathcal{L}(\theta)
  = \mathcal{L}_{\mathrm{data}}
  + \mathcal{L}_{\mathrm{pde}}
  + \mathcal{L}_{\mathrm{bc}},
  \label{eq:loss_total}
\end{equation}
with
\begin{equation}
  \mathcal{L}_{\mathrm{data}}
  = \frac{1}{N_d}\sum_{i=1}^{N_d}
  \bigl\|
  \boldsymbol{q}_\theta(\boldsymbol{x}_i^{d})
  - \boldsymbol{q}_i^{\mathrm{obs}}
  \bigr\|_2^2,
  \label{eq:loss_data}
\end{equation}
\begin{equation}
  \mathcal{L}_{\mathrm{pde}}
  = \frac{1}{N_f}\sum_{j=1}^{N_f}
  \bigl\|
  \boldsymbol{r}_\theta(\boldsymbol{x}_j^{f})
  \bigr\|_2^2,
  \qquad
  \mathcal{L}_{\mathrm{bc}}
  = \frac{1}{N_b}\sum_{k=1}^{N_b}
  \bigl\|
  \boldsymbol{r}^{\mathrm{bc}}_\theta(\boldsymbol{x}_k^{b})
  \bigr\|_2^2.
  \label{eq:loss_pde_bc}
\end{equation}
Minimizing $\mathcal{L}(\theta)$ yields a neural approximation
$\boldsymbol{q}_\theta$ that interpolates the sparse sensor data while
approximately satisfying the governing equations and boundary conditions
over the entire domain. This framework naturally accommodates both forward
problems, in which $\boldsymbol{\mu}$ is known and the goal is to reconstruct
$\boldsymbol{q}$, and inverse problems, in which $\boldsymbol{\mu}$ or other
latent parameters are learned jointly with $\theta$ by differentiating the
loss with respect to them.

In this work, the vanilla PINN formulation above provides the backbone of
our VSOPINN architecture: the decoder network outputs the flow
variables on a structured reference grid, while the physics residuals derived
from the Navier--Stokes equations are evaluated at collocation points and
enter the loss in the same manner as
$\mathcal{L}_{\mathrm{pde}}$ and $\mathcal{L}_{\mathrm{bc}}$.

\subsection{PINN with Voronoi enhanced sensor optimization (VSOPINN) }

\begin{figure}[htbp]
\centering
\includegraphics[width=1.0\textwidth]{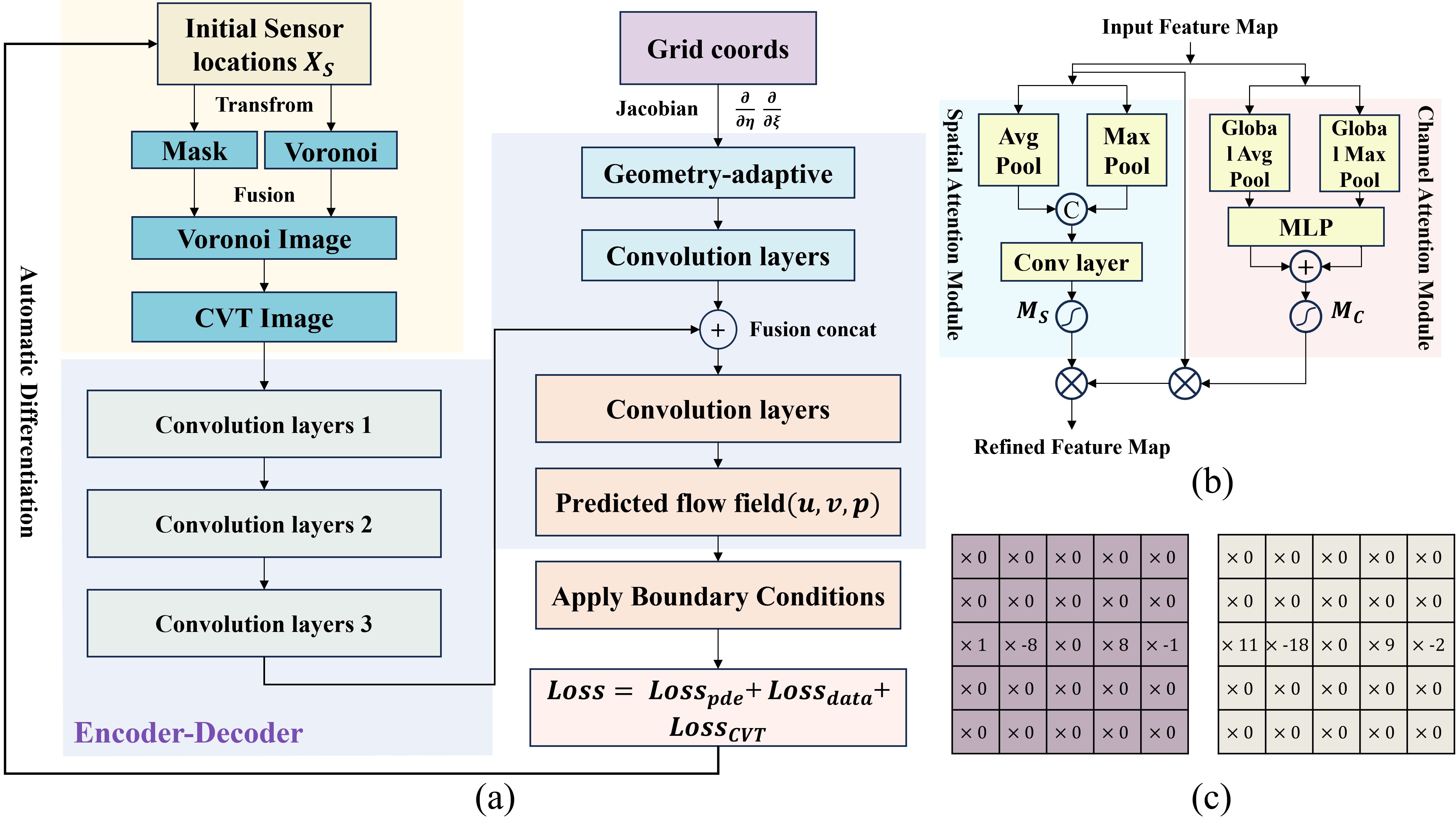}
\caption{Diagram of the VSOPINN model. (a) Encoder/Decoder modules, (b) Attention-enhanced convolutional decoder. (c) Finite difference convolution kernel.}
\label{fig:Architecture}
\end{figure}

Based on the framework of CNNs, this work proposes a novel flow field inversion architecture that couples geometric-adaptive coordinate transformation with discrete-point Voronoi regularization. As illustrated in Figure~\ref{fig:Architecture}, the proposed network consists of two main components: a Voronoi and CVT image generation branch, and an Encoder--Decoder branch consisting of the geometric-adaptive coordinate transformation module. 

The geometric-adaptive coordinate transformation branch maps the irregular physical domain into a regular reference domain through coordinate transformation. This operation enables the computation of partial derivatives on the transformed domain via conventional CNN operations, thereby facilitating the seamless integration of physical laws into the network. The Voronoi branch projects sparse sensor measurements onto a structured grid by generating a Voronoi diagram, which effectively converts unstructured pointwise data into image-like representations. This transformation leverages the intrinsic strength of CNNs in extracting spatial features from gridded data.

The feature maps extracted from both branches through a series of convolutional layers are subsequently concatenated. A channel attention mechanism could be applied to enhance the interaction and fusion between sensor-derived data and coordinate information. Furthermore, hard boundary conditions are embedded explicitly into the network to ensure physical consistency. Finally, the physical quantities in the true domain are recovered from the reference domain, and the corresponding physics-informed loss terms are formulated to impose governing equations.

This integrated approach allows end-to-end training with simultaneous optimization of sensor placement and field reconstruction, while maintaining adherence to the underlying physical constraints. Next, we will describe the two components in detail. 

\subsubsection{Voronoi image generation method from sparse sensors}
\label{sec:voronoi_image_generation}

In this section, we first introduce the image generation algorithm based on the Voronoi diagram, including the energy function and the optimization method.

\paragraph{Energy function}
Let $\Omega \subset \mathbb{R}^d$ be a bounded computational domain and  
$X = \{\boldsymbol{x}_1,\dots,\boldsymbol{x}_n\}$ a set of sites in $\Omega$.

The Voronoi cell associated with $\boldsymbol{x}_i$ is defined as
\begin{equation}
\Omega_i = \Bigl\{\boldsymbol{x}\in\Omega \;\Big|\;
\|\boldsymbol{x}-\boldsymbol{x}_i\| \le \|\boldsymbol{x}-\boldsymbol{x}_j\|,
\ \forall j\ne i \Bigr\},
\end{equation}
and the collection $\{\Omega_i\}_{i=1}^n$ forms a Voronoi tessellation of $\Omega$.

To control the spatial distribution of nodes, we endow $\Omega$ with a non-negative
density function $\rho(\boldsymbol{x})$. Following the variational
characterization of centroidal Voronoi tessellations (CVTs), we introduce the energy
\begin{equation}
E_1(X)
=
\sum_{i=1}^n
\int_{\Omega_i}
\rho(\boldsymbol{x})\,\|\boldsymbol{x}-\boldsymbol{x}_i\|^2\,\mathrm{d}\sigma .
\label{eq:cvt_energy}
\end{equation}

Minimizers of $E_1$ correspond to CVTs: at a stationary point, each site coincides
with the mass centroid of its Voronoi cell with respect to $\rho(\boldsymbol{x})$.
In the present context, $\rho(\boldsymbol{x})$ is constructed from the current
pointwise reconstruction error on the reference grid (see Eq.~\eqref{eq:rho_def}),
so that regions with larger error attract more sensors.

\paragraph{Optimization}
In our implementation, the CVT refinement is carried out in a discrete form on the
structured reference grid. At a CVT update step, we first build a non-negative
density map $\rho(\boldsymbol{z}_j)$ on the grid nodes $\{\boldsymbol{z}_j\}_{j=1}^{N_g}$
using the current pointwise reconstruction error (Eq.~\eqref{eq:rho_def}).
We then assign each grid node to its nearest sensor, yielding a hard Voronoi partition.
Each sensor is relocated to the density-weighted centroid of its associated cell, where
the integrals in Eq.~\eqref{eq:cvt_energy} are approximated by summations over grid nodes.
This density-weighted Lloyd update provides an efficient descent step for the CVT energy,
and it is applied periodically every $K_{\mathrm{CVT}}$ iterations during training
(Algorithm~\ref{alg:voronoi_cvt}).

\begin{algorithm}[t]
\caption{Voronoi rasterization with periodic CVT relocation}
\label{alg:voronoi_cvt}
\SetAlgoLined

\KwIn{Current iteration $\mathrm{iter}$; sensor locations $\mathbf{X}_s=\{\boldsymbol{s}_i\}_{i=1}^{N_s}\subset[0,1]^2$;
sensor measurements $\mathbf{U}=\{\boldsymbol{q}_i\}_{i=1}^{N_s}$;
reference-grid size $(H,W)$; sharpness $\alpha$; CVT interval $K_{\mathrm{CVT}}$; current grid fields $\hat{\boldsymbol{q}}$ and $\boldsymbol{q}^{\mathrm{ref}}$}
\KwOut{Voronoi image $\mathbf{I}_{\mathrm{vor}}$; CVT-refined image $\mathbf{I}_{\mathrm{cvt}}$; updated locations $\mathbf{X}_s^{+}$}

Compute soft Voronoi weights $\pi_{ij}$ on the grid using Eq.~\eqref{eq:soft_voronoi_weight}\;
Rasterize $\mathbf{I}_{\mathrm{vor}}$ by weighted interpolation of $\{\boldsymbol{q}_i\}$ and construct the soft mask\;

\eIf{$\mathrm{iter} \bmod K_{\mathrm{CVT}} \neq 0$}{
    $\mathbf{X}_s^{+} \gets \mathbf{X}_s$\;
    $\mathbf{I}_{\mathrm{cvt}} \gets \mathbf{I}_{\mathrm{vor}}$\;
}{
    Construct density $\rho(\boldsymbol{z}_j)$ from the pointwise reconstruction error using Eq.~\eqref{eq:rho_def}\;
    Compute hard Voronoi labels $\ell(j)=\arg\min_i \|\boldsymbol{z}_j-\boldsymbol{s}_i\|_2$\;
    Relocate sensors by density-weighted centroids of their cells and project back to $[0,1]^2$\;
    Set $\mathbf{X}_s^{+}$ to the relocated sensors and re-rasterize $\mathbf{I}_{\mathrm{cvt}}$\;
}
\Return{$\mathbf{I}_{\mathrm{vor}}, \mathbf{I}_{\mathrm{cvt}}, \mathbf{X}_s^{+}$}
\end{algorithm}uo

\subsubsection{Voronoi image Encoder and PDE solver Decoder modules}
\label{sec:enc_dec}

When the sensor locations remain fixed throughout the training process, the corresponding Voronoi diagram and Mask remain unchanged. In this case, they are generated via direct assignment without requiring differentiation. Specifically, the Voronoi diagram is constructed using nearest-neighbor interpolation: for each grid point, the closest sensor is identified, and the physical value of that sensor is assigned to the grid. The Mask is generated by assigning a value of 1 to grid points where a sensor is present and 0 elsewhere.
However, when sensor positions are subject to optimization, the entire process must remain differentiable to maintain a valid gradient chain for automatic differentiation. This ensures that sensor locations can be updated effectively and that both the Voronoi diagram and the Mask can be continuously adapted. The conventional nearest-neighbor assignment method leads to discontinuities in the gradient flow, preventing gradient-based sensor updates.
To address this issue, we introduce a differentiable construction of a soft Voronoi diagram and a soft Mask. Let $\boldsymbol{s}_i\in[0,1]^2$ denote the normalized sensor coordinates and
$\{\boldsymbol{z}_j\}_{j=1}^{N_g}$ the reference-grid nodes in pixel coordinates.
We map $\boldsymbol{s}_i$ to pixel coordinates
$\boldsymbol{x}_i=[(H-1)s_{i,y},\,(W-1)s_{i,x}]$ and define
$\mathrm{diag}^2=(H-1)^2+(W-1)^2$.
The normalized squared distance is
$d_{ij}^2=\|\boldsymbol{z}_j-\boldsymbol{x}_i\|_2^2/(\mathrm{diag}^2+\varepsilon)$.
The soft Voronoi assignment is
\begin{equation}
\pi_{ij}
=
\frac{\exp\!\left(-\alpha\, d_{ij}^2\right)}
{\sum_{k=1}^{N_s}\exp\!\left(-\alpha\, d_{kj}^2\right)},
\qquad \alpha>0,
\label{eq:soft_voronoi_weight}
\end{equation}
where $\alpha$ controls the sharpness of the partition.
Each image channel is rasterized by a weighted sum
$\mathbf{I}_{\mathrm{vor}}(\boldsymbol{z}_j)=\sum_{i=1}^{N_s}\pi_{ij}\,\boldsymbol{q}_i$.
In our implementation, $\boldsymbol{q}_i$ is obtained by differentiable bilinear sampling
(\texttt{grid\_sample}) of the reference fields at $\boldsymbol{s}_i$.
The soft Voronoi weights are Lipschitz continuous with respect to grid node coordinates and have bounded gradients, a property that ensures the stability of sensor placement optimization.
Meanwhile, the soft mask is defined by an exponential distance decay, which is piecewise differentiable
and works robustly in practice. This approach preserves differentiability throughout the assignment process, enabling end-to-end optimization of sensor placement while maintaining a continuous gradient path.

\noindent\textbf{Voronoi image Encoder.}
This module is designed to extract spatial features from sparse sensor data by encoding the CVT-refined Voronoi image into a grid-based feature tensor, which adapts the unstructured sensor measurements to the spatial feature extraction capability of CNNs.
After the CVT-guided sensor relocation described in Sec.~\ref{sec:voronoi_image_generation}, the sparse
measurements are rasterized on the reference grid and yield two image tensors:
the raw Voronoi image and the CVT-refined image. In the forward pass of the
network, we take the CVT image as input to the sensor encoder so that the
convolutional branch learns directly from the loss-adapted sensor
configuration. Let $\mathbf{I}_{\mathrm{cvt}} \in
\mathbb{R}^{H\times W\times C_{\mathrm{in}}}$ denote the multi-channel CVT
image, where the channels collect the interpolated flow variables and the
corresponding mask information. The Voronoi image Encoder is implemented as a
lightweight CNN
\begin{equation}
  \mathbf{F}_{\mathrm{V}}(\boldsymbol{\xi})
  =
  E_{\mathrm{V}}\bigl(\mathbf{I}_{\mathrm{cvt}};\,\theta_{\mathrm{V}}\bigr),
  \qquad
  \boldsymbol{\xi}\in\hat{\Omega},
\end{equation}
where $E_{\mathrm{V}}$ denotes the encoder mapping with trainable parameters
$\theta_{\mathrm{V}}$, $\hat{\Omega}$ is the structured reference domain, and
$\mathbf{F}_{\mathrm{V}}\in\mathbb{R}^{H\times W\times C_{\mathrm{V}}}$ is the
resulting feature tensor on the grid.
 The output tensor
$\mathbf{F}_{\mathrm{V}}$ is finally projected to have the same spatial size
and channel dimension as the feature map $\mathbf{F}_{\mathrm{G}}$ produced by
the geometry-adaptive branch, so that the two representations can be fused by
channel-wise concatenation,
\begin{equation}
  \mathbf{F}_{\mathrm{fuse}}
  =
  \mathrm{Concat}\bigl(\mathbf{F}_{\mathrm{V}},\,\mathbf{F}_{\mathrm{G}}\bigr)
  \in \mathbb{R}^{H\times W\times C_{\mathrm{fuse}}},
  \label{eq:fuse_features}
\end{equation}
which serves as the input to the subsequent decoder block
described below.

\noindent\textbf{PDE solver decoder.}
As the core module for flow field reconstruction, this decoder takes the fused feature tensor $\mathbf{F}_{\mathrm{fuse}}$  as input to generate high-fidelity flow field predictions that satisfy the Navier-Stokes equations and boundary conditions, through the reconstruction of features with embedding of physical constraints. 
While standard convolutional layers serve as the backbone for feature reconstruction, we introduce a lightweight attention mechanism designed as a plug-and-play module to further enhance representational power. 
To enhance feature representation, we incorporate a mechanism based on the Convolutional Block Attention Module (CBAM)~\cite{ferrari_cbam_2018}. This block sequentially computes attention maps across both channel and spatial dimensions, enabling the network to prioritize salient features while suppressing irrelevant information. Consequently, this process adaptively recalibrates the intermediate features, thereby significantly enhancing the network's capability to extract informative representations pertinent to both sparse sensor measurements and geometric mappings.

The geometry-adaptive branch parameterizes a smooth mapping
$\boldsymbol{T}:(\xi,\eta)\mapsto (x,y)$ from the structured reference domain
$\hat{\Omega}$ to the physical flow domain $\Omega$. The Jacobian matrix
$\boldsymbol{J}(\xi,\eta) = \partial(x,y)/\partial(\xi,\eta)$ is obtained by
automatic differentiation of $\boldsymbol{T}$ with respect to $(\xi,\eta)$ and is
used to convert derivatives on the reference grid into derivatives in physical
coordinates. For the predicted field $\hat{\mathbf{q}}$, first-order derivatives
with respect to $(x,y)$ are computed by
\begin{equation}
  \begin{bmatrix}
    \partial_x \hat{\mathbf{q}}\\[2pt]
    \partial_y \hat{\mathbf{q}}
  \end{bmatrix}
  =
  \boldsymbol{J}^{-T}
  \begin{bmatrix}
    \partial_\xi \hat{\mathbf{q}}\\[2pt]
    \partial_\eta \hat{\mathbf{q}}
  \end{bmatrix},
  \label{eq:geom_chainrule}
\end{equation}
where $\partial_\xi \hat{\mathbf{q}}$ and $\partial_\eta \hat{\mathbf{q}}$ are
approximated on the reference grid using finite-difference stencils. Higher-order
derivatives such as the Laplacian are assembled from repeated applications of {Eq.}\eqref{eq:geom_chainrule}. Substituting these derivatives into the governing
equations yields the
physics-residual field $\mathcal{R}(\boldsymbol{x};\theta,\mathbf{X}_s)$, which is
sampled at collocation points for the PINN loss. Dirichlet boundary conditions are
imposed by hard embedding, i.e.\ by replacing the network outputs at boundary nodes
with the prescribed values in physical space; Neumann-type conditions are incorporated
through residual terms involving the normal derivatives computed from $\boldsymbol{T}$.

The CVT-based term establishes a link between the decoder and the sensor layout.
At a CVT update step, we construct a non-negative density field $\rho(\boldsymbol{x})$
on the reference grid that reflects where the current reconstruction is most inaccurate.
In the present experiments, the density is defined as the pointwise $\ell_2$ mismatch
of the reconstructed flow variables,
\begin{equation}
\rho(\boldsymbol{z}_j)
=
\left\|
\hat{\boldsymbol{q}}(\boldsymbol{z}_j)-\boldsymbol{q}^{\mathrm{ref}}(\boldsymbol{z}_j)
\right\|_2
+\varepsilon,
\qquad \varepsilon>0,
\label{eq:rho_def}
\end{equation}
where $\hat{\boldsymbol{q}}=(\hat{u},\hat{v},\hat{p})$ is the network prediction and
$\boldsymbol{q}^{\mathrm{ref}}=(u,v,p)$ is the reference field on the grid.
Let $\{V_m\}_{m=1}^{N_s}$ denote the Voronoi cells associated with the sensor locations
$\mathbf{X}_s = \{\boldsymbol{x}_s^{(m)}\}_{m=1}^{N_s}$. The discrete CVT energy is
\begin{equation}
  \mathcal{L}_{\mathrm{CVT}}(\mathbf{X}_s)
  =
  \sum_{m=1}^{N_s}
  \sum_{\boldsymbol{x}\in V_m}
  \rho(\boldsymbol{x})
  \bigl\|\boldsymbol{x}-\boldsymbol{x}_s^{(m)}\bigr\|_2^{2}.
  \label{eq:cvt_loss}
\end{equation}

Accordingly, the VSOPINN objective is written as
\begin{equation}
\mathcal{L}_{\mathrm{VSOPINN}}
=
\lambda_{\mathrm{data}} L_{\mathrm{data}}+\lambda_{\mathrm{pde}} L_{\mathrm{pde}}
+
\lambda_{\mathrm{cvt}}\,\mathcal{L}_{\mathrm{CVT}}.
\label{eq:loss_vsopinn}
\end{equation}
We adopt an alternating optimization strategy.
During standard iterations, network parameters are updated by backpropagation using
$\mathcal{L}_{\mathrm{data}}+\mathcal{L}_{\mathrm{pde}}$.
Every $K_{\mathrm{CVT}}$ iterations (set to 500 in our experiments), the density field is refreshed
using Eq.~\eqref{eq:rho_def}, and sensor locations are relocated by one density-weighted CVT step
(Algorithm~\ref{alg:voronoi_cvt}), which approximately minimizes $\mathcal{L}_{\mathrm{CVT}}$
with respect to $\mathbf{X}_s$.

\subsubsection{PINN with Voronoi enhanced sensor optimization under multi-conditions (multi-VSOPINN)}
\label{sec:multiSOPINN}

\begin{figure}[htbp]
\centering
\includegraphics[width=1.0\textwidth]{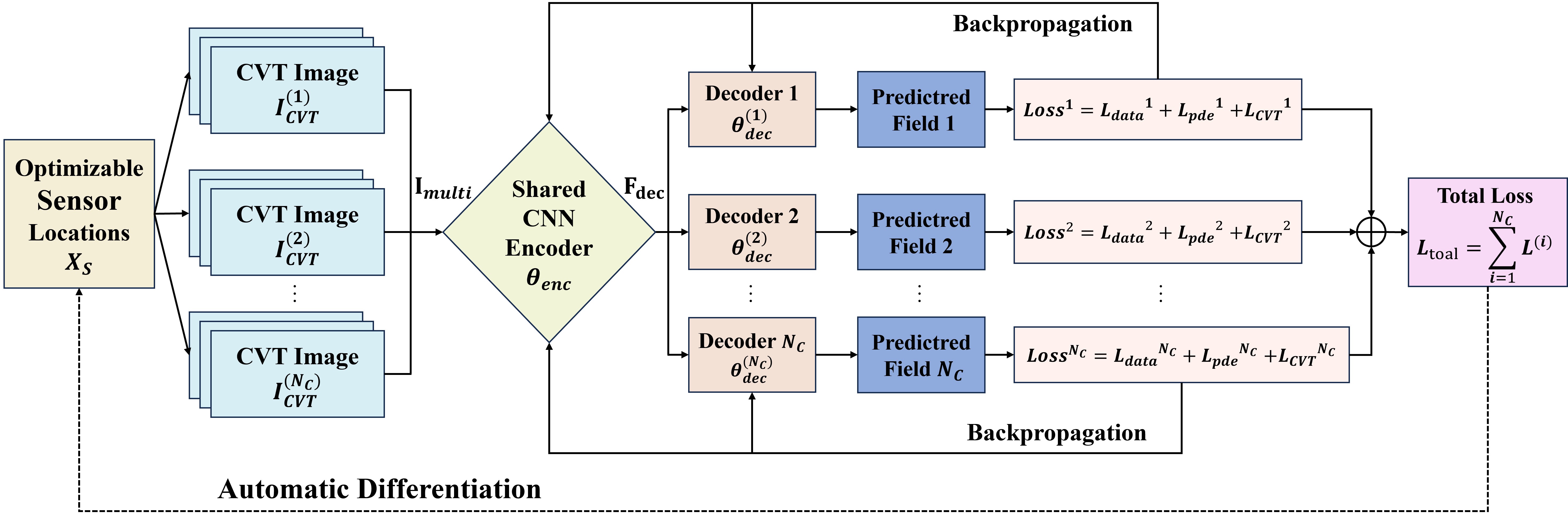}
\caption{Single-encoder/multi-decoder architecture of the multi-VSOPINN framework.}
\label{fig:Architecture_multi}
\end{figure}

In many applications the flow field must be reconstructed under a family of
operating conditions, such as different boundary conditions, Reynolds numbers,
or geometric configurations. We denote these conditions by
$\{\mathcal{C}_m\}_{m=1}^{N_c}$. Our objective is to determine a single sensor layout $\mathbf{X}_s$ that is universally effective across all operating conditions $\mathcal{C}_m$, effectively capturing the diverse dynamics of each case.

To this end, we adopt a single-encoder/multi-decoder design, as schematically illustrated in Fig.~\ref{fig:Architecture_multi}. The Voronoi branch, the geometry-adaptive branch, and the fusion block
together form a shared encoder that is common to all operating conditions. For
each condition $\mathcal{C}_m$, sparse measurements are first rasterized into a
CVT-refined Voronoi image $\mathbf{I}_{\mathrm{cvt}}^{(m)}$ on the reference
grid using the procedure in Sec.~\ref{sec:voronoi_image_generation}. These images are then stacked along the
channel dimension to build a multi-condition input
\begin{equation}
  \mathbf{I}_{\mathrm{multi}}
  =
  \mathrm{Concat}\bigl(
    \mathbf{I}_{\mathrm{cvt}}^{(1)},\ldots,
    \mathbf{I}_{\mathrm{cvt}}^{(N_c)}
  \bigr)
  \in \mathbb{R}^{H\times W\times C_{\mathrm{multi}}}.
\end{equation}
The shared encoder $E_{\mathrm{shared}}(\cdot;\theta_{\mathrm{enc}})$ maps
$\mathbf{I}_{\mathrm{multi}}$ to a latent feature tensor
\begin{equation}
  \mathbf{F}_{\mathrm{dec}}
  =
  E_{\mathrm{shared}}\bigl(
    \mathbf{I}_{\mathrm{multi}};\theta_{\mathrm{enc}}
  \bigr)
  \in \mathbb{R}^{H\times W\times C_{\mathrm{dec}}},
\end{equation}
which aggregates information from all operating conditions in a common
representation. In particular, by stacking tensors from different $\mathcal{C}_m$, the shared encoder facilitates the extraction of universal physical features into $\mathbf{F}_{\mathrm{dec}}$. This design enables the model to exploit structural similarities across varying conditions, allowing the reconstruction of each specific case to benefit from the latent information inherently shared among them.

On top of this shared encoder we attach a set of condition-specific PDE
decoders $\{D_{\mathrm{pde}}^{(m)}(\cdot;\theta_{\mathrm{pde}}^{(m)})\}_{m=1}^{N_c}$.
Each decoder is responsible for one operating condition and receives the same
latent tensor $\mathbf{F}_{\mathrm{dec}}$, but uses its own
geometry-adaptive mapping and output layer to account for the corresponding
PDE operator and parameters.

Training is performed by minimizing a weighted sum of data and PDE losses over
all operating conditions, together with the CVT regularization defined in
Eq.~\eqref{eq:cvt_loss}. Let $    \mathcal{L}_{\mathrm{data}}^{(m)}$ and
$\mathcal{L}_{\mathrm{pde}}^{(m)}$ denote the data misfit and physics residual loss for
condition $\mathcal{C}_m$, computed in the same way as in the single-condition
case. Training is performed by minimizing the sum of data and PDE losses over all operating
conditions, together with the periodic CVT regularization on the shared sensor set
$\mathbf{X}_s$. The multi-condition objective reads
\begin{equation}
  \mathcal{L}_{\mathrm{multi}}
  =
  \sum_{m=1}^{N_c}
  \Bigl(
    \mathcal{L}_{\mathrm{data}}^{(m)}
    +
    \mathcal{L}_{\mathrm{pde}}^{(m)}
  \Bigr)
  +
  \lambda_{\mathrm{cvt}}\,\mathcal{L}_{\mathrm{CVT}}(\mathbf{X}_s).
  \label{eq:loss_multi}
\end{equation}
In our experiments, all conditions are equally weighted through the direct summation above.
The CVT relocation is applied every $K_{\mathrm{CVT}}$ iterations to update the shared sensor
layout. For the CVT density in the multi-condition setting, we construct $\rho(\boldsymbol{x})$
from a representative condition; specifically, we use the error field of the last condition
in our ordering (typically the highest Reynolds number), which empirically yields a robust
layout for the entire condition set.
\section{Results}
\label{sec:results}

This section presents four systematic numerical experiments to comprehensively validate the proposed VSOPINN framework, focusing on its dual core capabilities: high-fidelity flow field reconstruction and adaptive sensor placement optimization.
The experimental design follows a progressive logic from basic to advanced scenarios: 
First, the benchmark lid-driven cavity flow is used to verify the framework’s fundamental reconstruction performance in regular geometric domains. 
Subsequently, a two-dimensional irregular vascular flow problem is introduced to test its geometric adaptability and the effectiveness of sensor optimization in unstructured domains with complex boundaries. 
Third, the annulus rotational flow case demonstrates the method’s competence in inverse modeling under rotational constraints, while further validating its ability to optimize sensor layouts in curved domains.
Finally, a multi-condition lid-driven cavity scenario is designed to assess the generalization capability of the proposed single-Encoder/multi-Decoder architecture across varying Reynolds numbers, leveraging CVT-guided sensor placement and attention-enhanced feature fusion to ensure robustness under diverse operating conditions.

Across all numerical examples, neural network weights are initialized using the Kaiming uniform method, with the Rectified Linear Unit (ReLU) serving as the activation function. 
The specific network architectures, including depth and channel width, are summarized in Table \ref{tab:network_parameters}. 
Global optimization is performed using the Adam optimizer. To ensure robust convergence, a dual learning rate strategy is employed: a base rate of $1 \times 10^{-3}$ for network parameters and an elevated rate of $1 \times 10^{-2}$ for learnable sensor coordinates to accelerate spatial exploration. 
All algorithms are implemented within the PyTorch framework and executed on a workstation equipped with an Intel Core i7-14650HX CPU and an NVIDIA GeForce RTX 4060 GPU. 
Unless otherwise specified, we use unit weights for the data and PDE residual terms in the training objective, i.e., $\lambda_{\mathrm{data}}=\lambda_{\mathrm{pde}}=1$.
For the multi-condition setting, per-condition losses are summed with equal weights, the sharpness parameter $\alpha$ is set to 80, and the CVT relocation is performed every $K_{\mathrm{CVT}}=500$ iterations.
Reconstruction accuracy is quantified using the relative $L^2$ error norm, complemented by qualitative visual comparisons between predicted and reference flow fields.

\begin{table}[htbp]
    \centering
    \caption{The network architecture configurations for the proposed VSOPINN framework in each case study, including depth, channel width, and kernel size.}
    \label{tab:network_parameters}
    \begin{tabular*}{\textwidth}{@{\extracolsep{\fill}}lccc@{}}
        \toprule
        \textbf{Problems} & \textbf{Depth} & \textbf{Width} & \textbf{Kernel Size ($K$)} \\
        \midrule
        Section \ref{sec:lid_driven} (Cavity) & 5 & 32, 64 & 5 \\
        Section \ref{sec:vascular} (Vascular) & 8 & 16, 32, 64 & 5 \\
        Section \ref{sec:annulus} (Annulus) & 8 & 16, 32, 64 & 5 \\
        Section \ref{sec:multi_condition} (Multi-cond) & 5 & 32, 64 & 5 \\
        \bottomrule
    \end{tabular*}
\end{table}

\subsection{Lid-driven cavity flow}
\label{sec:lid_driven}

To validate the efficacy of the proposed network architecture, we first conduct a verification on the two-dimensional lid-driven cavity flow, a benchmark problem characterized by a regular geometric structure but rich flow physics.
The governing laws for this steady-state incompressible flow are the Navier-Stokes equations, expressed as:
\begin{equation}
    (\mathbf{v} \cdot \nabla) \mathbf{v} = -\nabla p + \frac{1}{Re} \nabla^2 \mathbf{v}, \quad \nabla \cdot \mathbf{v} = 0,
    \label{eq:ns_cavity}
\end{equation}
where $\mathbf{v} = (u, v)$ denotes the velocity vector field, $p$ represents the pressure, and $Re$ is the Reynolds number, defined as $Re = UL/\nu$ with characteristic velocity $U$, characteristic length $L$, and kinematic viscosity $\nu$. In this case study, the Reynolds number is set to $Re=100$.

The flow is simulated within a unit square domain $\Omega = [0,1] \times [0,1]$. Consistent with the standard benchmark configuration, the boundary conditions are prescribed as follows: the top boundary ($y=1$) is driven with a constant tangential velocity $\mathbf{v} = (1, 0)$, while the no-slip condition $\mathbf{v} = (0, 0)$ is imposed on the bottom ($y=0$), left ($x=0$), and right ($x=1$) walls. Physically, the motion of the top lid induces a shear force that drives the fluid, creating a large primary vortex centered within the cavity. Due to the conservation of momentum and the confined geometry, the pressure field exhibits significant gradients; specifically, singularities occur at the top corners where the moving lid meets the stationary walls. The pressure is relatively high at the top-right corner (stagnation region) where the fluid impinges on the wall, and drops significantly towards the vortex center, maintaining the rotational motion.

To systematically evaluate the contribution of each module in the VSOPINN framework, we conducted a comprehensive ablation study using six distinct network configurations:
(1) {PhyGeoNet}: The baseline model incorporating only the geometry-adaptive mechanism without sensor data;
(2) {PhyGeoNet + Data}: The baseline model augmented with sparse sensor observations ($N_s=4$);
(3) {PhyGeoNet + Data + Voronoi}: The model integrating sensor data with a fixed Voronoi tessellation to assist feature extraction;
(4) {VSOPINN(Sensor Opt)}: The framework where sensor positions are learnable via gradient descent;
(5) {VSOPINN(with Attention)}: The framework augmented with a attention enhanced convolutional decoder;
(6) {VSOPINN(with CVT)}: The complete framework employing Centroidal Voronoi Tessellation strategies for robust sensor optimization.

All models were trained for 30,000 iterations to ensure convergence. The reconstruction accuracy is quantified by the relative $L^2$ error of the velocity magnitude as summarized in Table \ref{tab:cavity_errors}. 
Quantitative results show a clear progressive improvement: the baseline PhyGeoNet achieves a high relative
error of $4.89 \times 10^{-1}$, while integrating sensor data and Voronoi tessellation reduces the error to $2.66 \times 10^{-1}$ by providing effective structural guidance.

Most notably, the activation of active sensor optimization (Model 4) further halves the error to $1.37 \times 10^{-1}$. The addition of the Attention mechanism (Model 5) provides marginal gains, refining the error to $1.25 \times 10^{-1}$. However, the VSOPINN method utilizing CVT optimization (Model 6) achieves the lowest relative error of $3.89 \times 10^{-2}$. This drastic reduction indicates that the CVT-optimized sensor layout effectively captures the critical flow features, such as the high-gradient regions near the singularities, which are often missed by static or locally optimized arrangements.

\begin{figure}[htbp]
    \centering
    \includegraphics[width=\textwidth]{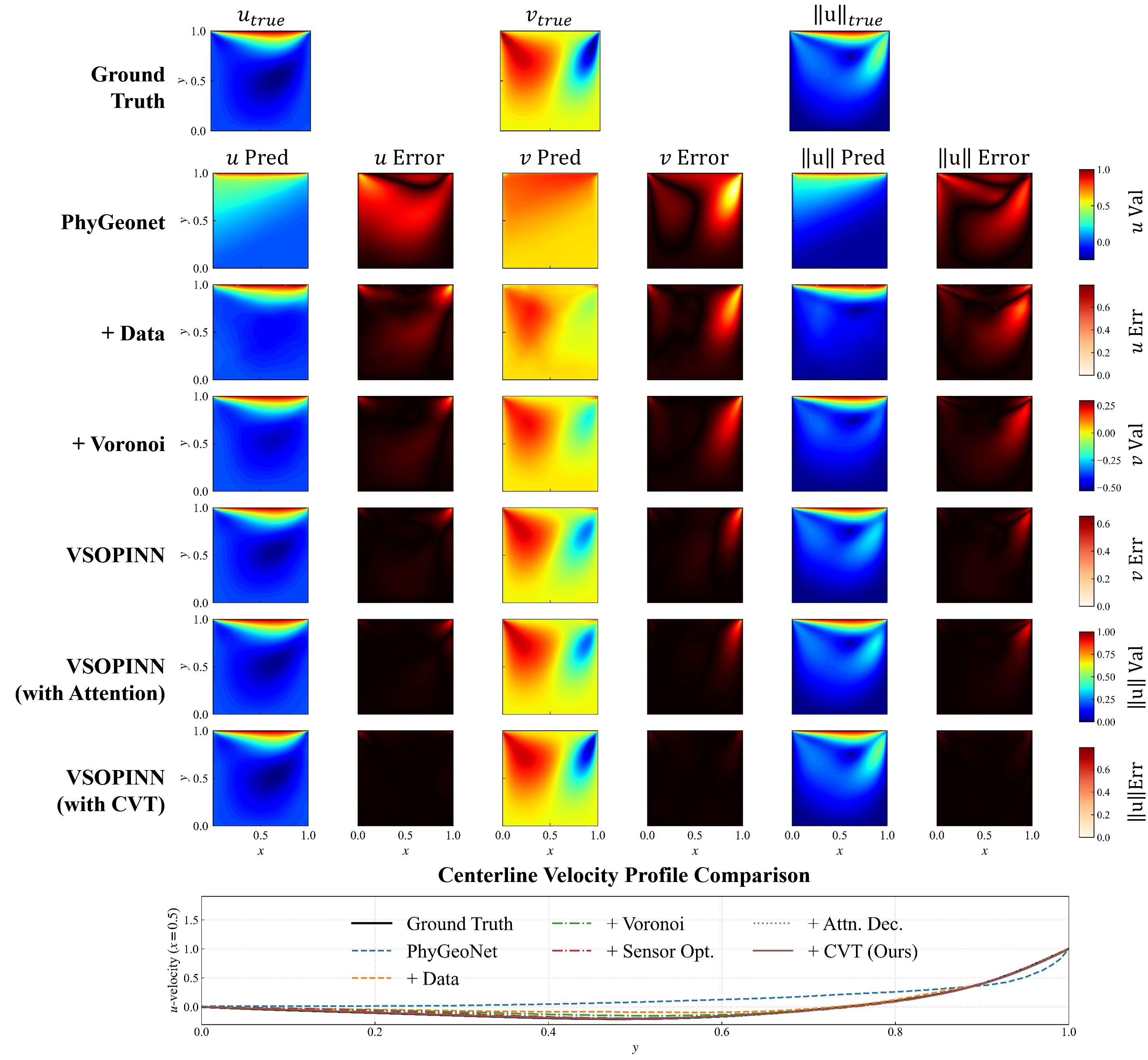} 
    \caption{Reconstructed velocity magnitude and pressure fields for the lid-driven cavity flow ($Re=100$). Compared with the CFD reference, the CVT-optimized sensor locations in VSOPINN yield more accurate resolutions of the primary vortex structures and corner singularities.}
    \label{fig:cavity_comparison}
\end{figure}

Figure \ref{fig:cavity_comparison} visualizes the reconstructed flow fields compared against the CFD reference. It can be visually observed that the VSOPINN (with CVT) predictions align most closely with the ground truth, particularly in capturing the curvature of the primary vortex and the pressure distribution near the boundaries, whereas the baseline methods exhibit slight deviations in the vortex core location and magnitude.

\begin{table}[htbp]
    \centering
    \caption{Ablation study of relative $L^2$ errors for the lid-driven cavity flow ($Re=100$) with $N_s=4$ sensors. The comparison includes velocity components ($u, v$) and velocity magnitude ($|\mathbf{v}|$). The integration of CVT optimization yields the most significant performance gain.}
    \label{tab:cavity_errors}
    \begin{tabular*}{\textwidth}{@{\extracolsep{\fill}}lccc@{}}
        \toprule
        \multirow{2}{*}{\textbf{Method Configuration}} & \multicolumn{3}{c}{\textbf{Rel. $L^2$ Error}} \\
        \cmidrule(l){2-4}
        & $\boldsymbol{u}$ & $\boldsymbol{v}$ & $|\mathbf{v}|$ \\
        \midrule
        PhyGeoNet (Origin) & 4.426e-01 & 8.602e-01 & 4.885e-01 \\
        PhyGeoNet + Data & 2.988e-01 & 6.665e-01 & 3.660e-01 \\
        PhyGeoNet + Data + Voronoi & 2.140e-01 & 4.884e-01 & 2.663e-01 \\
        VSOPINN (Sensor Opt) & 1.128e-01 & 2.745e-01 & 1.366e-01 \\
        VSOPINN (with Attention) & 9.826e-02 & 2.500e-01 & 1.252e-01 \\
        \textbf{VSOPINN (with CVT)} & \textbf{4.258e-02} & \textbf{5.973e-02} & \textbf{3.892e-02} \\
        \bottomrule
    \end{tabular*}
\end{table}

\begin{figure}[htbp]
    \centering
    \includegraphics[width=0.8\textwidth]{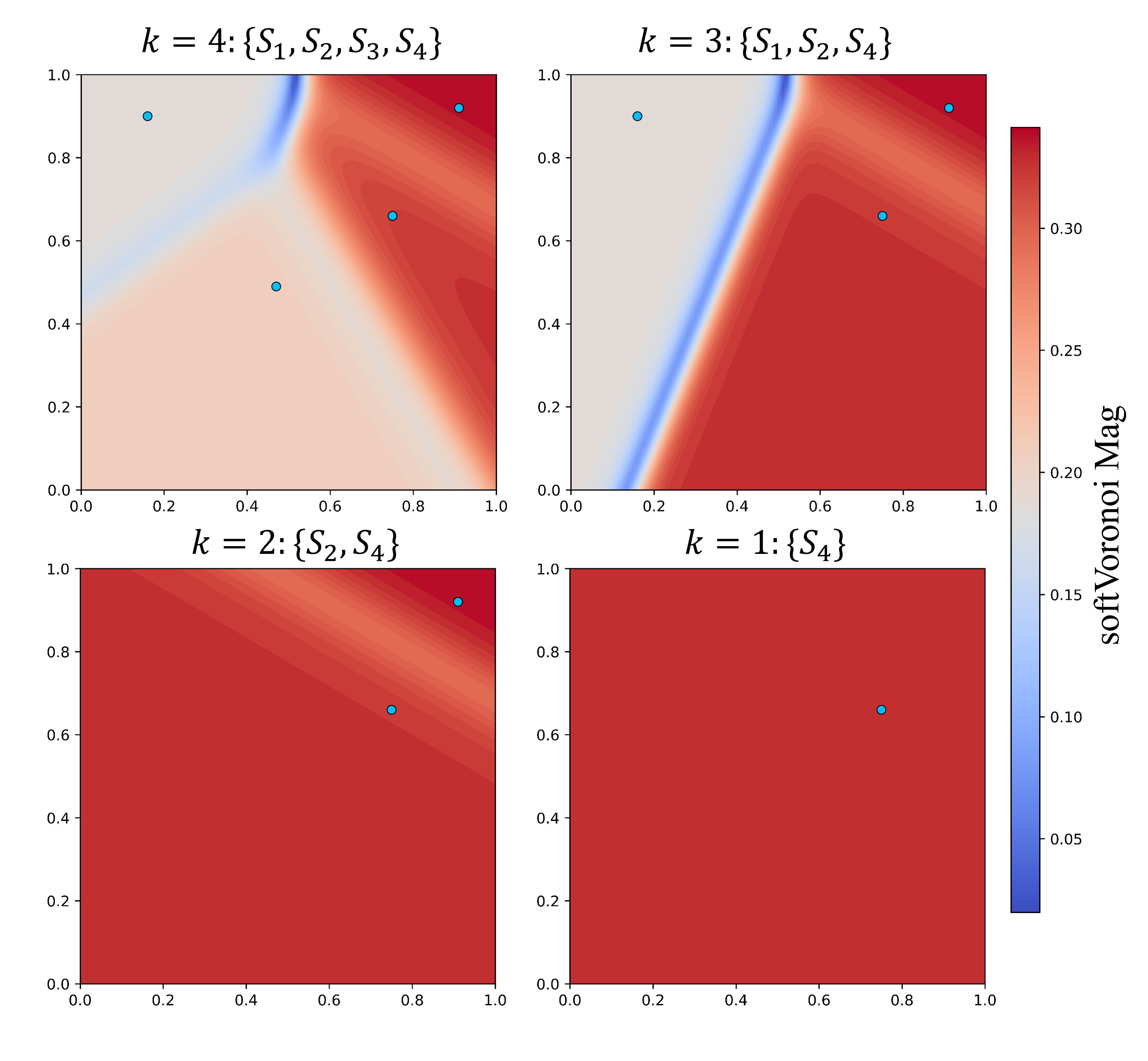}
    \caption{Soft-Voronoi images of the velocity magnitude $|\mathbf{v}|$ for the lid-driven cavity flow ($Re=100$), constructed using the best-performing sensor subsets at different numbers of active sensors: $k=4$ ($\{S_1,S_2,S_3,S_4\}$), $k=3$ ($\{S_1,S_2,S_4\}$), $k=2$ ($\{S_2,S_4\}$), and $k=1$ ($\{S_4\}$), as highlighted in Table~\ref{tab:cavity_sensor_failure}. The rasterization follows the differentiable soft assignment in Eq.~\eqref{eq:soft_voronoi_weight}, providing an image-like representation of sparse pointwise measurements on the reference grid.}
    \label{fig:cavity_softvoronoi}
\end{figure}

To further evaluate robustness under sensor failures, we perform a post-training inference study based on the CVT-optimized four-sensor layout obtained in the complete VSOPINN framework, denoted as $\{S_1,S_2,S_3,S_4\}$. 
Without any retraining or fine-tuning, we mimic partial outages by activating only a subset of these sensors ($k=1$--$3$) and conducting a single forward pass of the trained model. 
Figure~\ref{fig:cavity_softvoronoi} further illustrates the differentiable soft Voronoi rasterization used in VSOPINN.
Specifically, we visualize the soft Voronoi images of the velocity magnitude $|\mathbf{v}|$ constructed from the best-performing sensor subsets for $k=4,3,2,$ and $1$ (highlighted in Table~\ref{tab:cavity_sensor_failure}).
As the number of active sensors decreases, each remaining sensor exhibits a broader influence region on the reference grid under the soft assignment in Eq.~\eqref{eq:soft_voronoi_weight}, which provides an intuitive explanation for the gradual (rather than catastrophic) performance degradation observed under moderate sensor loss.
Table~\ref{tab:cavity_sensor_failure} reports the resulting relative $L^2$ errors for $(u,v)$ and the velocity magnitude $|\mathbf{v}|$.
As expected, the reconstruction accuracy decreases when fewer sensors remain available, yet the degradation is gradual. 
For example, the nominal four-sensor case achieves $\epsilon_{|\mathbf{v}|}=4.214\times10^{-2}$, while the best three-sensor subset $\{S_1,S_2,S_4\}$ remains close at $5.526\times10^{-2}$; even with only two sensors $\{S_2,S_4\}$, the error stays at $8.141\times10^{-2}$. 
In contrast, single-sensor inference can be markedly less reliable (e.g., $\{S_3\}$ yields $\epsilon_{|\mathbf{v}|}=2.505\times10^{-1}$), suggesting that robustness depends not only on sensor count but also on the information content of the surviving locations. 
Overall, these results indicate that the learned representation generalizes beyond a fixed sensor set and maintains usable reconstructions under moderate sensor loss.

\begin{table}[htbp]
    \centering
    \caption{Robustness test for the lid-driven cavity flow ($Re=100$) under sensor failures. The VSOPINN(with CVT) model is fixed after training with the CVT-optimized four-sensor set $\{S_1,S_2,S_3,S_4\}$, and then evaluated using all sensor subsets via a single forward pass.}
    \label{tab:cavity_sensor_failure}
    \begin{tabular*}{\textwidth}{@{\extracolsep{\fill}}clccc@{}}
        \toprule
        \textbf{$k$} & \textbf{Active sensors} & $\epsilon_u$ & $\epsilon_v$ & $\epsilon_{|\mathbf{v}|}$ \\
        \midrule
        \textbf{4} & \textbf{$\{S_{1},S_{2},S_{3},S_{4}\}$} & \textbf{4.707e-02} & \textbf{6.828e-02} & \textbf{4.214e-02} \\
        \addlinespace[2pt]
        \midrule
        \addlinespace[2pt]
        \textbf{3} & \textbf{$\{S_{1},S_{2},S_{4}\}$} & \textbf{6.843e-02} & \textbf{1.154e-01} & \textbf{5.526e-02} \\
        3 & $\{S_{2},S_{3},S_{4}\}$ & 1.028e-01 & 8.850e-02 & 8.032e-02 \\
        3 & $\{S_{1},S_{2},S_{3}\}$ & 1.004e-01 & 1.218e-01 & 8.661e-02 \\
        3 & $\{S_{1},S_{3},S_{4}\}$ & 1.011e-01 & 2.382e-01 & 1.481e-01 \\
        \addlinespace[2pt]
        \midrule
        \addlinespace[2pt]
        \textbf{2} & \textbf{$\{S_{2},S_{4}\}$} & \textbf{1.006e-01} & \textbf{1.341e-01} & \textbf{8.141e-02} \\
        2 & $\{S_{2},S_{3}\}$ & 1.346e-01 & 1.345e-01 & 1.089e-01 \\
        2 & $\{S_{1},S_{2}\}$ & 1.454e-01 & 2.068e-01 & 1.184e-01 \\
        2 & $\{S_{1},S_{4}\}$ & 1.075e-01 & 2.415e-01 & 1.441e-01 \\
        2 & $\{S_{3},S_{4}\}$ & 1.432e-01 & 2.450e-01 & 1.668e-01 \\
        2 & $\{S_{1},S_{3}\}$ & 1.765e-01 & 3.645e-01 & 2.181e-01 \\
        \addlinespace[2pt]
        \midrule
        \addlinespace[2pt]
        \textbf{1} & \textbf{$\{S_{4}\}$} & \textbf{1.182e-01} & \textbf{2.451e-01} & \textbf{1.475e-01} \\
        1 & $\{S_{2}\}$ & 2.221e-01 & 2.398e-01 & 1.508e-01 \\
        1 & $\{S_{1}\}$ & 9.850e-02 & 3.027e-01 & 1.637e-01 \\
        1 & $\{S_{3}\}$ & 2.374e-01 & 3.695e-01 & 2.505e-01 \\
        \addlinespace[2pt]
        \midrule
        \addlinespace[2pt]
        \textbf{0} & \textbf{$\varnothing$} & \textbf{4.426e-01} & \textbf{8.602e-01} & \textbf{4.885e-01} \\
        \bottomrule
    \end{tabular*}
\end{table}

\subsection{Vascular flow in irregular domains}
\label{sec:vascular}

Following the benchmark validation, we investigate a two-dimensional vascular flow problem to evaluate the framework's geometric adaptability and robustness within unstructured domains. The geometric configuration represents a typical irregular biological vessel with bifurcation, where the complex boundaries challenge standard convolutional operations. The governing laws remain the steady-state Navier-Stokes equations as defined in Eq. (\ref{eq:ns_cavity}).

Consistent with the reference setup, a constant inflow velocity profile $\mathbf{v} = (0, 1)$ is imposed at the inlet, while a stress-free condition ($\nabla \mathbf{v} \cdot \mathbf{n} = 0, p=0$) is applied at the outlet. No-slip conditions are enforced on all vessel walls. This case presents a significant challenge due to the complex irregular boundaries, where flow separation and acceleration occur near the bifurcation point. To rigorously test the method's performance under strong non-linearity, we focus on a high-Reynolds-number regime ($Re=450$).

We assessed the reconstruction capability using five network configurations. Table \ref{tab:vascular_errors} summarizes the relative reconstruction errors. The high non-linearity at $Re=450$ renders the baseline reconstruction challenging ($\epsilon_{|\mathbf{v}|} \approx 1.19 \times 10^{-1}$). While the geometry-adaptive backbone (PhyGeoNet) effectively handles the irregular domain boundaries, the integration of optimized sensor data in the VSOPINN framework further constrains the solution space. The standard sensor optimization reduces the error to $1.03 \times 10^{-1}$. However, the implementation of CVT optimization achieves the lowest error of $9.67 \times 10^{-2}$.

\begin{figure}[htbp]
    \centering
    \includegraphics[width=\textwidth]{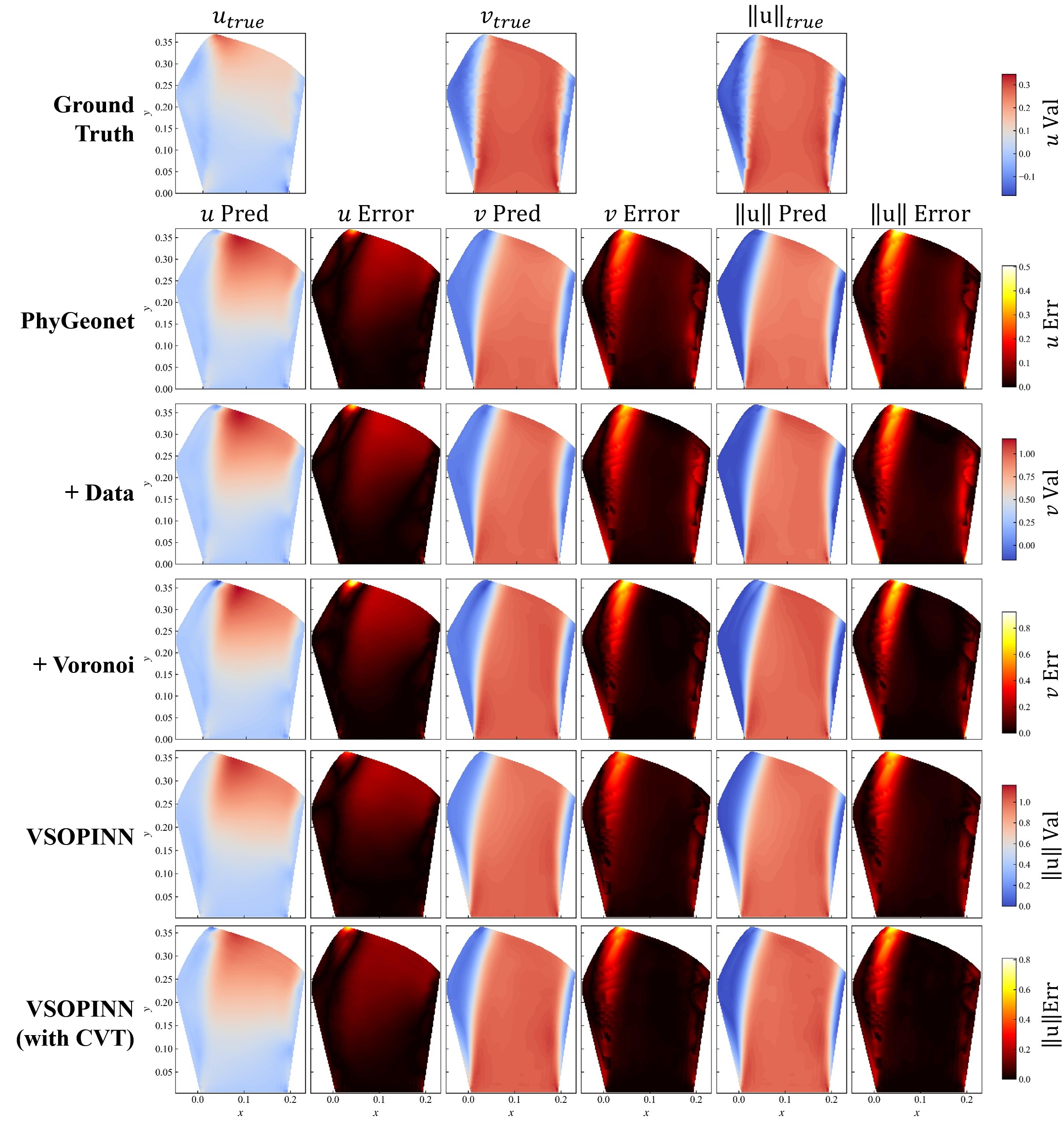} 
    \caption{Flow field and absolute error distributions for the vascular flow ($Re=450$). VSOPINN exhibits superior accuracy over the baseline reference, particularly in capturing flow features near the bifurcation and unstructured wall boundaries.}
    \label{fig:vascular_flow}
\end{figure}

The qualitative comparison is presented in Figure \ref{fig:vascular_flow}, which visualizes the velocity magnitude and error distributions for the $Re=450$ case. It can be observed that the baseline methods tend to smooth out local variations near the bifurcation point where flow separation occurs. In contrast, the VSOPINN method maintains high fidelity in these critical regions. This demonstrates that the synergy between the geometry-adaptive convolution and the Voronoi-integrated CVT optimization enables the network to effectively mitigate the difficulties posed by complex geometries and high-nonlinear dynamics.

\begin{table}[htbp]
    \centering
    \caption{Quantitative comparison of the relative $L^2$ error of velocity magnitude ($|\mathbf{v}|$) for vascular flow reconstruction at $Re=450$ ($N_s=4$).}
    \label{tab:vascular_errors}
    \begin{tabular}{lc}
        \toprule
        \textbf{Method} & \textbf{Rel. $L^2$ Error ($|\mathbf{v}|$)} \\
        \midrule
        PhyGeoNet (Origin) & 1.192e-01 \\
        PhyGeoNet + Data & 1.188e-01 \\
        PhyGeoNet + Data + Voronoi & 1.099e-01 \\
        VSOPINN (Sensor Opt) & 1.031e-01 \\
        \textbf{VSOPINN (with CVT)} & \textbf{9.678e-02} \\
        \bottomrule
    \end{tabular}
\end{table}

\subsection{Rotational flow in annulus geometry}
\label{sec:annulus}

In the third case study, we challenge the framework with a rotational flow problem constrained by curved boundaries. The computational domain consists of an annulus defined by an inner radius $R_{in}$ and an outer radius $R_{out}$. Distinct from scalar field diffusion, this setup is defined as a vector field fluid dynamics problem where the network must resolve the non-linear coupling of velocity components induced by rotational driving forces.

Mathematically, the boundary conditions are imposed as Dirichlet constraints on the velocity components in the Cartesian coordinate system: a no-slip condition is enforced on the stationary outer boundary ($r = R_{out}$), while the velocity on the rotating inner boundary ($r = R_{in}$) is determined by the constant angular velocity $\Omega$. We investigate a high-speed rotation case with $\Omega = 14$ rad/s ($Re \approx 350$). This rigorous constraint tests the method's capability to accurately resolve the shear stress gradients derived from the coordinate-transformed boundary velocities.

Experiments were conducted to evaluate the sensor optimization dynamics. Unlike standard convolutional filters which operate on rectilinear grids and struggle to approximate curvature, the VSOPINN framework adapts the sensor locations to the flow physics. Table \ref{tab:annulus_errors} presents the reconstruction accuracy. Consistent with previous findings, the pure physics baseline yields a high error of $6.39 \times 10^{-1}$. The introduction of CVT optimization drastically reduces this to $6.89 \times 10^{-2}$.

\begin{figure}[htbp]
    \centering
    \includegraphics[width=\textwidth]{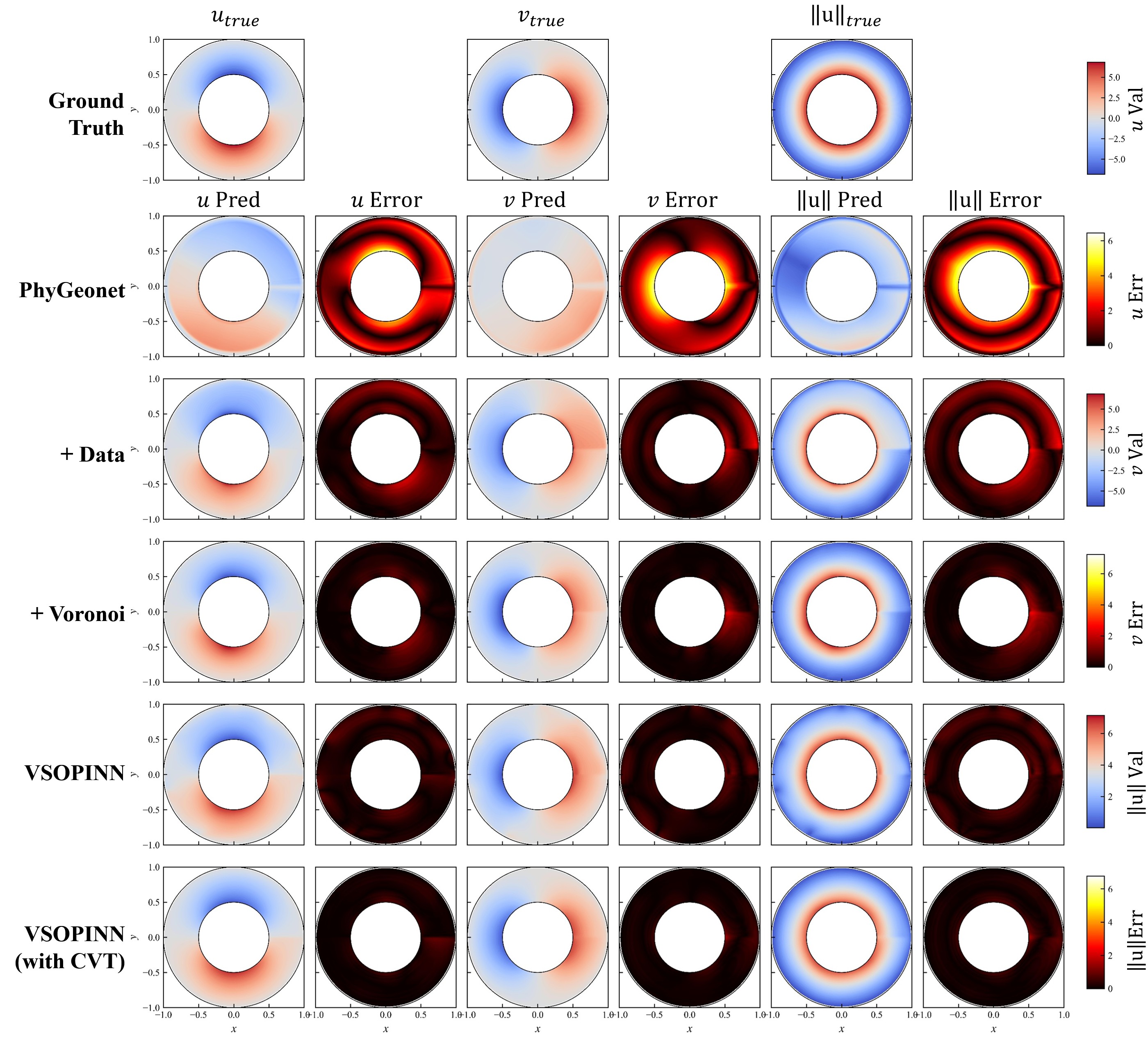} 
    \caption{Reconstructed flow fields for the annulus problem ($\Omega=14$). Comparisons among the five network models and the ground truth highlight their respective capabilities in resolving nonlinear velocity distributions and steep shear stress gradients near the high-speed rotating inner boundary.}
    \label{fig:annulus_optimization}
\end{figure}

Figure \ref{fig:annulus_optimization} provides a visual insight into the optimization process. It can be observed that the learnable sensors tend to migrate from their initial random distributions towards the inner rotating boundary and regions of high velocity curvature. This autonomous migration indicates that the VSOPINN framework successfully identifies the regions with the highest information entropy—specifically, the boundary layer where the rotational momentum is transferred to the fluid. By concentrating observations in these critical zones, the network effectively compensates for the discretization errors inherent in projecting curved geometries onto structured grids.

\begin{table}[htbp]
    \centering
    \caption{Relative $L^2$ errors for the annulus driving flow under high-speed rotation ($\Omega=14$) with optimized sensor placement.}
    \label{tab:annulus_errors}
    \begin{tabular*}{\textwidth}{@{\extracolsep{\fill}}lccc@{}}
        \toprule
        \multirow{2}{*}{\textbf{Method Configuration}} & \multicolumn{3}{c}{\textbf{Rel. $L^2$ Error}} \\
        \cmidrule(l){2-4}
        & $\boldsymbol{u}$ & $\boldsymbol{v}$ & $|\mathbf{v}|$ \\
        \midrule
        PhyGeoNet (Origin) & 6.698e-01 & 7.664e-01 & 6.396e-01 \\
        PhyGeoNet + Data & 2.048e-01 & 2.202e-01 & 2.089e-01 \\
        PhyGeoNet + Data + Voronoi & 1.255e-01 & 1.509e-01 & 1.319e-01 \\
        VSOPINN (Sensor Opt) & 9.921e-02 & 1.151e-01 & 1.017e-01 \\
        \textbf{VSOPINN (with CVT)} & \textbf{6.808e-02} & \textbf{7.869e-02} & \textbf{6.885e-02} \\
        \bottomrule
    \end{tabular*}
\end{table}

\begin{figure}[htbp]
    \centering
    \includegraphics[width=\textwidth]{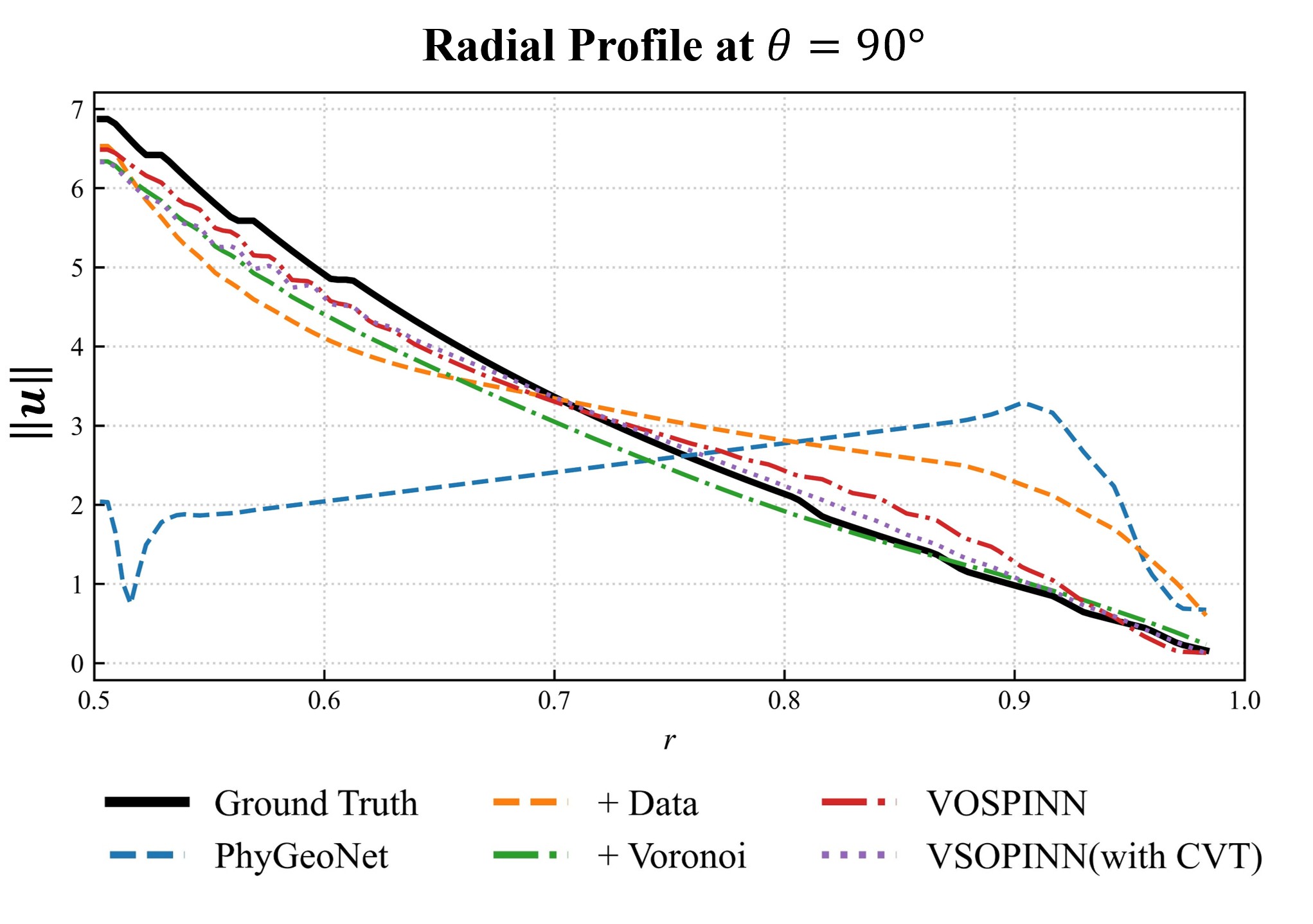} 
    \caption{Radial velocity magnitude profiles at $\theta=90^\circ$, demonstrating the varying accuracy of the models in capturing the steep velocity gradients near the boundaries.}
    \label{fig:annulus_line}
\end{figure}

\subsection{Multi-condition lid-driven cavity flow}
\label{sec:multi_condition}

To further evaluate the generalization capability and robustness of the proposed framework under varying operating conditions, we extended the investigation to a multi-condition scenario. Unlike the single-case validation in Section \ref{sec:lid_driven}, this experiment challenges the network to learn a unified representation of flow physics across a range of Reynolds numbers ($Re \in \{100, 300, 400, 800\}$), where flow structures undergo significant variations. For this purpose, we employed the Single-Encoder/Multi-Decoder architecture, which allows the model to share a common latent feature space while decoding distinct flow fields.

\begin{figure}[htbp]
    \centering
    \includegraphics[width=0.6\textwidth]{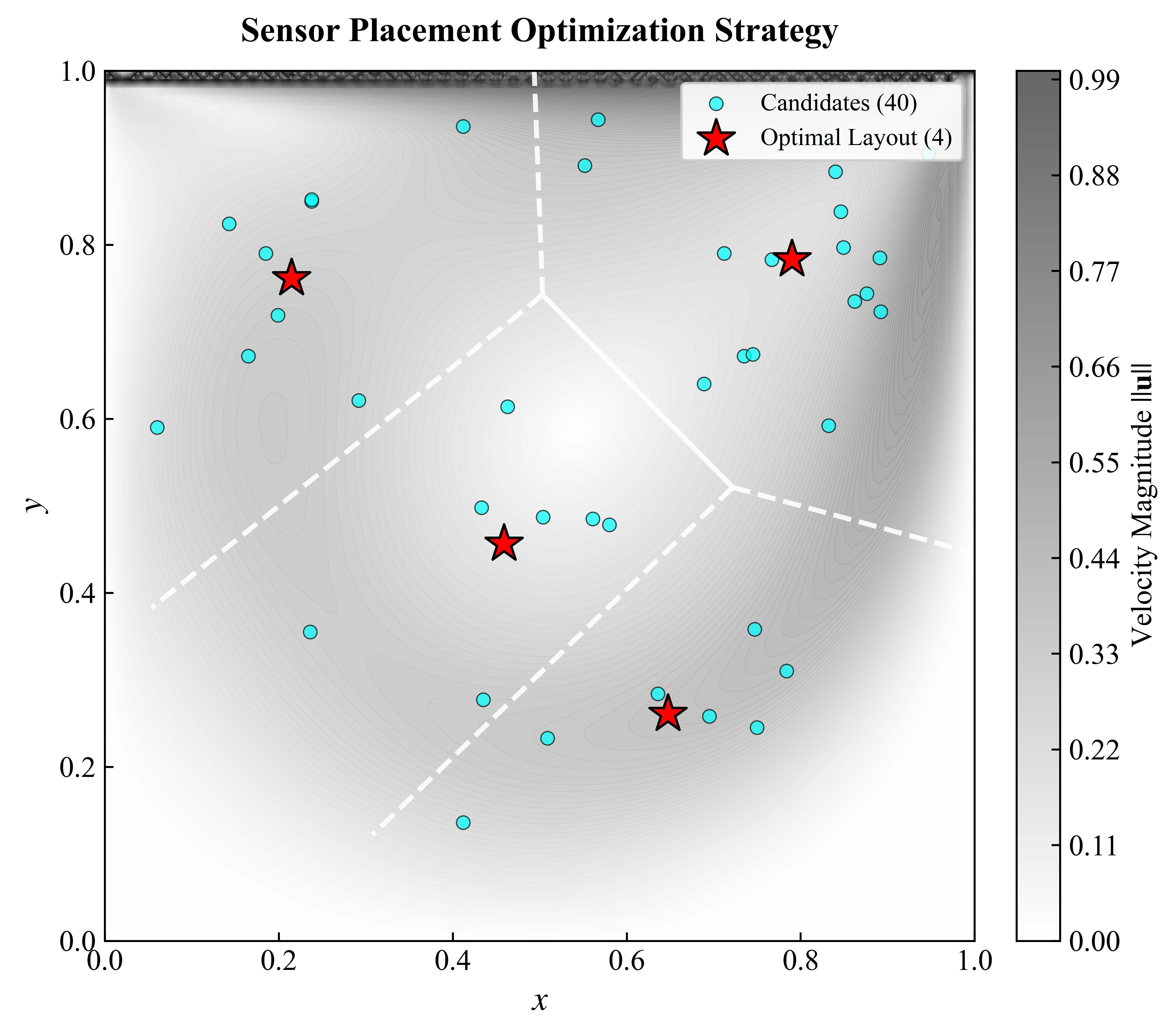}
    \caption{Visualization of the sensor placement strategy. The background displays the velocity magnitude field at $Re=600$ (interpolation case). Cyan dots represent the 40 candidate sensor positions generated from 10 independent training sessions, which tend to cluster in high-velocity regions. Red stars indicate the final Optimal Layout derived via k-means clustering, partitioned by the Voronoi tessellation (white dashed lines).}
    \label{fig:sensor_optimization}
\end{figure}

A critical challenge in this setting is determining a sensor layout that remains effective across all conditions. To address this, we implemented a robust sensor placement strategy based on ensemble optimization and k-means clustering. Specifically, ten independent training sessions were conducted, generating a pool of 40 candidate sensor positions. As visualized in Figure \ref{fig:sensor_optimization}, these positions (cyan dots) naturally cluster around regions of high flow gradients and the primary vortex, indicating high information entropy common to the trained regimes. Subsequently, k-means clustering was applied to partition these candidates into four clusters. The resulting centroids (red stars in Figure \ref{fig:sensor_optimization}) were selected as the final Optimal Layout, with the domain partitioned by the corresponding Voronoi tessellation to ensure coverage.

The generalization performance of this optimal layout was rigorously tested against unseen data, including an interpolation case at $Re=600$ and an extrapolation case at $Re=1000$. For comparative analysis, a baseline configuration consisting of fixed, randomly distributed sensors (``Random Layout'') was established. Table \ref{tab:multi_condition_errors} quantitatively compares the reconstruction accuracy. It is evident that the layout derived via k-means clustering consistently yields lower relative $L^2$ errors. In the extrapolation regime ($Re=1000$), where non-linearity is most pronounced, the Optimal Layout achieves an error of $2.03 \times 10^{-1}$ compared to $2.62 \times 10^{-1}$ for the random baseline.

\begin{figure}[htbp]
    \centering
    \includegraphics[width=\textwidth]{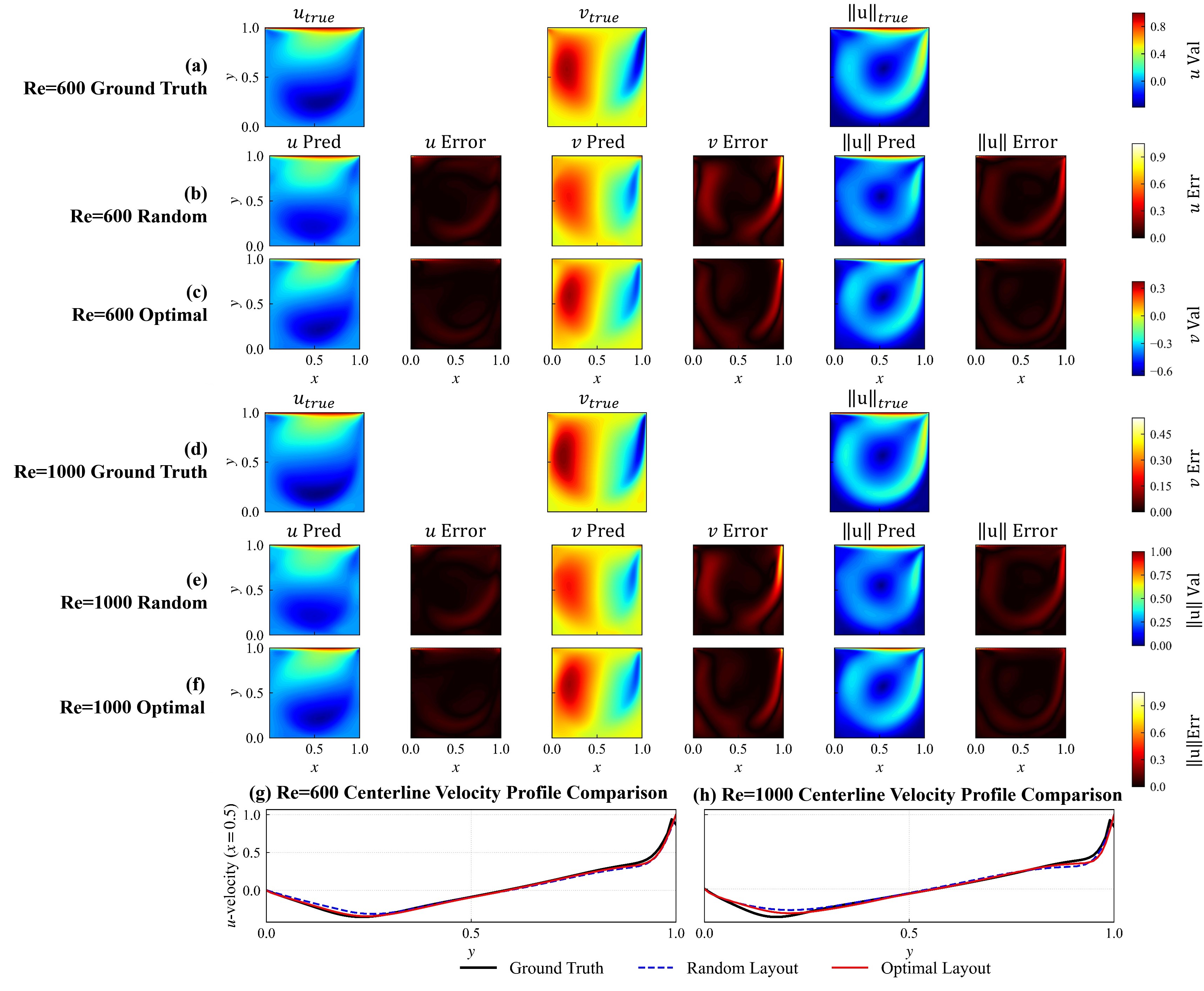} 
    \caption{Comparative visualization of flow field reconstruction across interpolation ($Re=600$) and extrapolation ($Re=1000$) regimes. Rows (a) and (d) show the Ground Truth. The Optimal Layout (Rows c, f) significantly mitigates local errors compared to the Random Layout (Rows b, e), particularly in the extrapolation regime. Subplots (g) and (h) show the centerline velocity profiles ($u$ at $x=0.5$), confirming the superior alignment of the proposed method with the reference solution.}
    \label{fig:multi_condition_vis}
\end{figure}

Figure \ref{fig:multi_condition_vis} presents the visual comparison of the reconstructed flow fields. The Random Layout baseline struggles to resolve the fine-scale structures of the primary vortex in high Re cases, resulting in higher local errors (see Row b and e). In contrast, the proposed method utilizing the Optimal Layout maintains high fidelity to the Ground Truth. This demonstrates that the sensor positions identified through the k-means strategy effectively support the network's inference capability even for flow conditions outside the training envelope. This is further corroborated by the centerline velocity profiles shown in Figure \ref{fig:multi_condition_vis}(g)-(h), where the Optimal Layout (red solid line) aligns significantly better with the Ground Truth than the Random Layout (blue dashed line).

\begin{table}[htbp]
    \centering
    \caption{Quantitative comparison of relative $L^2$ errors for flow reconstruction under interpolation ($Re=600$) and extrapolation ($Re=1000$) conditions. The comparison encompasses velocity components ($u, v$), velocity magnitude ($|\mathbf{v}|$), and pressure ($p$).}
    \label{tab:multi_condition_errors}
    \resizebox{\textwidth}{!}{%
        \begin{tabular}{lcccccccc}
            \toprule
            \multirow{2}{*}{\textbf{Sensor Layout}} & \multicolumn{4}{c}{\textbf{Interpolation ($Re=600$)}} & \multicolumn{4}{c}{\textbf{Extrapolation ($Re=1000$)}} \\
            \cmidrule(lr){2-5} \cmidrule(lr){6-9}
            & $\epsilon_u$ & $\epsilon_v$ & $\epsilon_{|\mathbf{v}|}$ & $\epsilon_p$ & $\epsilon_u$ & $\epsilon_v$ & $\epsilon_{|\mathbf{v}|}$ & $\epsilon_p$ \\
            \midrule
            Random (Fixed) & 1.729e-01 & 2.030e-01 & 1.650e-01 & 3.221e-01 & 2.417e-01 & 3.331e-01 & 2.621e-01 & 3.433e-01 \\
            \textbf{Optimal (k-means)} & \textbf{1.514e-01} & \textbf{1.434e-01} & \textbf{1.313e-01} & \textbf{2.842e-01} & \textbf{2.226e-01} & \textbf{2.186e-01} & \textbf{2.034e-01} & \textbf{2.880e-01} \\
            \bottomrule
        \end{tabular}%
    }
\end{table}

\section{Discussion}
\label{sec:discussion}

In this study, we introduce the VSOPINN framework to address the persistent challenge of reconstructing high-fidelity physical fields from sparse, unstructured sensor data, particularly within geometrically complex domains. By integrating Voronoi tessellation with geometry-adaptive convolutional networks, our approach effectively circumvents the topological constraints of standard CNNs, which intrinsically require rectilinear grid inputs. This integration functions not merely as a discretization tool but as a structural bridge that preserves the spatial correlation between irregular sensor observations and the governing physics. The empirical evidence from the benchmark lid-driven cavity flow corroborates this mechanism; the incorporation of Voronoi-assisted features alone yielded a relative $L^2$ error reduction of approximately 45\% compared to the baseline geometry-adaptive network. This significant improvement suggests that explicitly encoding the sensor topology provides the neural surrogate with a robust inductive bias, enabling it to infer global field structures even when observational data is severely limited.

A critical innovation of the proposed framework is its capacity to resolve boundary layer dynamics in non-rectilinear geometries, a regime where traditional grid-based deep learning methods often falter due to staircase approximations of curved boundaries. The annulus driving flow experiment (Section \ref{sec:annulus}) offers compelling validation of this capability. In this scenario, the rigid enforcement of rotational velocities creates steep shear stress gradients that are notoriously difficult to capture with standard convolutional filters. Consequently, the physics-only baseline struggled to converge, resulting in a high relative error of $6.39 \times 10^{-1}$. In contrast, by enabling autonomous sensor migration via Centroidal Voronoi Tessellation (CVT) optimization, the proposed method achieved a drastic error reduction to $6.89 \times 10^{-2}$. Visual analysis reveals that the learnable sensors spontaneously clustered near the inner rotating boundary ($r=R_{in}$) and the vascular bifurcation points (Section \ref{sec:vascular}). This phenomenon aligns with information theory principles, indicating that the network autonomously identifies and prioritizes regions of maximum information entropy—specifically, the boundary layers where vorticity generation and momentum transfer are most active—thereby compensating for the discretization errors inherent in coordinate transformations.

Furthermore, the deployment of the Single-Encoder/Multi-Decoder architecture in multi-condition scenarios demonstrates the framework's superior generalization capability. While convolutional layers excel at extracting local features, the non-local nature of fluid dynamics implies that downstream flow structures are path-dependent on upstream conditions. The integration of the Attention mechanism addresses this by capturing global spatiotemporal dependencies, effectively expanding the receptive field beyond the limits of local kernels. This architecture proved particularly advantageous in the extrapolation regime ($Re=1000$), where the flow exhibits stronger non-linearity than the training cases. Quantitative results confirm that the sensor layout derived via k-means clustering outperformed the random baseline, reducing the reconstruction error $\epsilon_{|\mathbf{v}|}$ from $2.62 \times 10^{-1}$ to $2.03 \times 10^{-1}$. This finding underscores the robustness of the k-means strategy, which effectively filters out condition-specific noise to identify a set of anchor observation points that remain physically informative across a wide range of Reynolds numbers.

Despite these advancements, it is important to acknowledge that the dynamic generation of Voronoi diagrams entails a computational overhead during the training phase. While the dual-learning rate strategy effectively accelerates sensor convergence, the repetitive tessellation process may impose scalability challenges when extending to three-dimensional problems with massive sensor arrays. Future avenues of research will therefore focus on optimizing the 3D implementation of VSOPINN, potentially through gradient-based sampling strategies or approximate Voronoi decompositions to mitigate computational costs. Moreover, we envision extending this sensor optimization paradigm to active learning settings, where the framework could dynamically guide mobile sensors in real-time experimental environments, further closing the loop between data-driven modeling and physical measurements.

\section*{CRediT authorship contribution statement}
\textbf{Renjie Xiao:} Methodology, Software, Validation, Visualization, Writing-original draft.
\textbf{Bingteng Sun:} Methodology, Data curation, Software, Writing - review \& editing.
\textbf{Yiling Chen:} Data curation, Supervision, Writing - review \& editing.
\textbf{Lin Lu:} Data curation, Supervision, Writing - review \& editing.
\textbf{Qiang Du:} Conceptualization, Supervision, Writing - original draft, Writing-review \& editing.
\textbf{Junqiang Zhu:} Conceptualization, Supervision, Writing - original draft, Writing-review \& editing.

\section*{Declaration of competing interest}
The authors declare that they have no known competing financial interests or personal relationships that could have
appeared to influence the work reported in this paper.

\section*{Data availability}
Data will be made available on request.

\section*{Acknowledgements}
The authors wish to acknowledge the project supported by Young Scientists Fund (A Class, Grant No. 52525603) and the financial support of Excellence Research Group Program (ERGP, the 
former Basic Science Center Program) (Grant No.52488101), the Strategic Priority Research Program of the Chinese Academy of Sciences, Grant No. XDA/B/C 0000000, and the cloud computing supported by the Beĳing Super Cloud Computing Center. Meanwhile, the current work is also supported by the Taishan Scholars Program.


\bibliographystyle{elsarticle-num}
\bibliography{PPL-ref}

\end{document}